\documentclass{mn2e} 
\usepackage{graphics, graphicx} 

\title[New insight into the circular polarisation of radio
pulsars.]  {$|V|$: New insight into the circular polarisation of radio
pulsars.}

\author[Karastergiou et al.]  {A.~Karastergiou$^{1,2}$,
S.~Johnston$^1$, D.~Mitra$^2$, A.~G.~J.~van~Leeuwen$^3$, and \and
R.~T.~Edwards$^4$\\
$^1$School of Physics, University of Sydney, NSW 2006, Australia\\
$^2$Max-Planck Institut f\"ur Radioastronomie, Auf dem H\"ugel 69,
53121 Bonn, Germany\\ 
$^3$Astronomical Institute, Utrecht University, Postbus 80000, 3508 TA
Utrecht, The Netherlands\\
$^4$Astronomical Institute ``Anton Pannekoek'', University of
Amsterdam, Kruislaan 403, 1098 SJ Amsterdam, The Netherlands\\
}

\date{Released 2003 Xxxxx XX}
 
\pagerange{\pageref{firstpage}--\pageref{lastpage}} \pubyear{2002} 
 
\def\LaTeX{L\kern-.36em\raise.3ex\hbox{a}\kern-.15em 
    T\kern-.1667em\lower.7ex\hbox{E}\kern-.125emX} 
 
\begin{document} 
 
\label{firstpage} 
 
\maketitle 
 
\begin{abstract} 
We present a study of single pulses from nine bright northern pulsars
to investigate the behaviour of circular polarisation, $V$. The
observations were conducted with the Effelsberg 100-m radio telescope
at 1.41 GHz and 4.85 GHz and the Westerbork radio telescope at 352
MHz. For the first time, we present the average profile of the
absolute circular polarisation $|V|$ in the single pulses. We
demonstrate that the average profile of $|V|$ is the distinguishing
feature between pulse components that exhibit low $V$ in the single
pulses and components that exhibit high $V$ of either handedness,
despite both cases resulting in a low mean. We also show that the
$|V|$ average profile remains virtually constant with frequency, which
is not generally the case for $V$, leading us to the conclusion that
$|V|$ is a key quantity in the pulsar emission problem.

\end{abstract} 
 

\section{Circular polarisation in pulsars}

Radio emission from pulsars is highly polarised. Occasionally the
linear polarisation $L$ is extraordinarily high, sometimes close to
$100\%$ of the total radio emission, and the circular polarisation $V$
is amongst the highest observed in natural sources of electromagnetic
radiation. Typically, however, it is only a small fraction of the
total pulsar emission, although in some cases it can be much larger
(i.e.$\sim 60\%$ in PSR B$1702-03$ in Radhakrishnan \& Rankin
1990\nocite{rr90}). The polarisation of pulsar emission has been
attributed to the emission mechanism itself and to propagation effects
in the pulsar magnetosphere (e.g. Melrose
2000\nocite{m00}). Simultaneous, multi-frequency observations in full
polarisation have shown irregularities in the polarisation of single
pulses at different frequencies, which have been considered direct
observational evidence of propagation effects(Karastergiou et
al. 2001, 2002; Karastergiou, Johnston \& Kramer 2003).

Average pulse profiles of radio pulsars are known to stabilise after
the integration of a sufficient number of pulses. This also holds for
the polarisation profile, although the stabilising time-scales for
polarisation are longer than for total power (Rathnasree \& Rankin
1995\nocite{rr95}). Inspecting such average profiles, Rankin (1983a,
b, 1986\nocite{ran83}\nocite{ran83a}\nocite{ran86}) claimed that there
is a difference in the circular polarisation in central or {\it core}
components from that in outer, {\it cone} components, thought to
originate in cones of emission around the magnetic axis. More
specifically, core components exhibit a higher degree of $V$ in the
integrated profile than cone components, often characterised by an
``S''-shaped swing from one to the other handedness (Clark \& Smith
1969\nocite{cs69}). On the other hand, cone components are weakly
circularly polarised in comparison to their core counterparts. In
fact, it was such differences, combined with a difference in the
spectral behaviour of core and cone components, that led Rankin to
suggest that there is a fundamental difference in the emission
mechanism between core and cone components.

Despite the efforts, it remains unclear whether $V$ is intrinsic to
the emission mechanism, propagation generated, or even both. Cordes,
Rankin \& Backer (1978) discovered that in PSR B2020+28, the
handedness of circular polarisation was associated with the
polarisation position angle of the linear polarisation. In individual
pulses from this pulsar, where the position angle occasionally jumps
abruptly by $90^o$ at certain pulse longitudes, these jumps where
accompanied by a sense reversal in $V$. The idea that $V$ is tied to
the orthogonal polarisation mode (OPM) phenomenon in pulsars was taken
further by Stinebring et al. (1984a, b) and more recently McKinnon \&
Stinebring (2000\nocite{ms00}) and McKinnon (2002\nocite{mck02}). An
investigation of the same issue in PSR B1133+16 using simultaneously
observed single pulses at 1.41 GHz and 4.85 GHz, demonstrated that the
$V$ - OPM association can be seen at both these frequencies, it is
however stronger at the lowest of the two (Karastergiou et al. 2003).

In many cases, the longitude-resolved distributions of $V$ in the
single pulses are much broader than the instrumental noise, despite
having a mean very close to zero. It can be easily inferred that such
broad $V$ distributions with an almost zero mean, denote a number of
highly circularly polarised single pulses. Integrated profiles of
pulsars with a very low mean $V$, however, obscure the information
that single pulses may be individually highly circularly polarised.

In this letter we present high quality single-pulse data from 9
pulsars, observed at 3 widely spaced frequencies. We investigate the
behaviour of $V$, by defining $|V|$ in the single pulses and looking
at differences between $V$ and $|V|$ in components that have been
classified in the literature as of core and cone origin. We also take
advantage of the three observed frequencies and discuss the frequency
evolution of $|V|$ as compared to $V$.
\section{The data}
\subsection{The observations}

The data presented here were obtained with the 100-m radio telescope
in Effelsberg and the Westerbork Synthesis Radio Telescope. They
consist of single pulse sequences from nine bright northern sky
pulsars. The Effelsberg observations were made at 4.85 GHz and 1.41
GHz between July and September 2002. Details of the receivers and the
calibration procedures of the multiplying polarimeters used can be
found in von Hoensbroech \& Xilouris (1997\nocite{hx97}). In short,
both systems have an equivalent system flux density of $\approx20$
Jy. For the 4.85 GHz observations a bandwidth of 500 MHz was used,
imposing restrictions on the resolution due to dispersion smearing. At
1.41 GHz, where dispersion smearing is significantly greater,
bandwidths of 10, 20 and 40 MHz were used depending on the dispersion
measure of the pulsar.

The 352 MHz observations, 30 minutes per pulsar, were conducted with
the WSRT on March 8th and 9th, 2003. The WSRT consists of fourteen
25-m dishes in an east-west array, which can be combined to form a
94-m single dish equivalent. The equivalent flux density of the system
was $\approx120$ Jy. We applied the schemes described by Weiler
(1973\nocite{weil73}) and Edwards et al. (2003\nocite{es03}) to
calibrate the data. Bandwidths of 10 MHz were processed by the backend
system PuMa (Vo{\^ u}te et al. 2002\nocite{vkh+02}), acting as a
digital filter-bank. The data were 4-bit sampled at between 0.2048 and
0.8192 ms, for between 32 and 1024 channels, depending on the pulsar
period and dispersion measure, and de-dispersed offline.

\subsection{Absolute circular polarisation}

We define the quantity $|V|$ in the single pulses as the absolute
value of $V$ in every phase bin of every individual pulse
(Karastergiou et al. 2001\nocite{khk+01}). By taking the absolute
value of the $V$ data stream, we are also altering the statistics of
the instrumental noise, which is no longer normally distributed and
has a non-zero mean. We subtract this positive mean from every single
pulse and integrate $|V|$ in all the pulses to produce an average
$|V|$ profile. Subtracting the aforementioned offset from the single
pulses has the result of an $|V|$ profile which can be closer to zero
than the profile of $V$. This happens in cases of low $|V|$ in the
single pulses, where subtracting the offset has the effect of removing
part of the signal. Full details of the data analysis and the
statistics of $|V|$ are outlined in Karastergiou et al. (2003b\nocite{kjm+03}).

\section{Results}
The nine pulsars that we observed display a variety of features in
both their total power profiles and their polarimetric properties. Our
single-pulse observations at three, widely-spaced frequencies give us
the opportunity to trace the properties of $V$ in the single pulses as
a function of frequency. Figs. 1, 2 and 3 show the integrated pulse
profiles of the observed pulsars, together with the profiles of $V$
and $|V|$. Positive and negative values of $V$ denote left and right
handed circular polarisation respectively. The alignment of the
profiles at the three frequencies observed is only approximate.
\begin{figure*}
\centerline{
\resizebox{\hsize}{!}{\includegraphics[angle=-90]{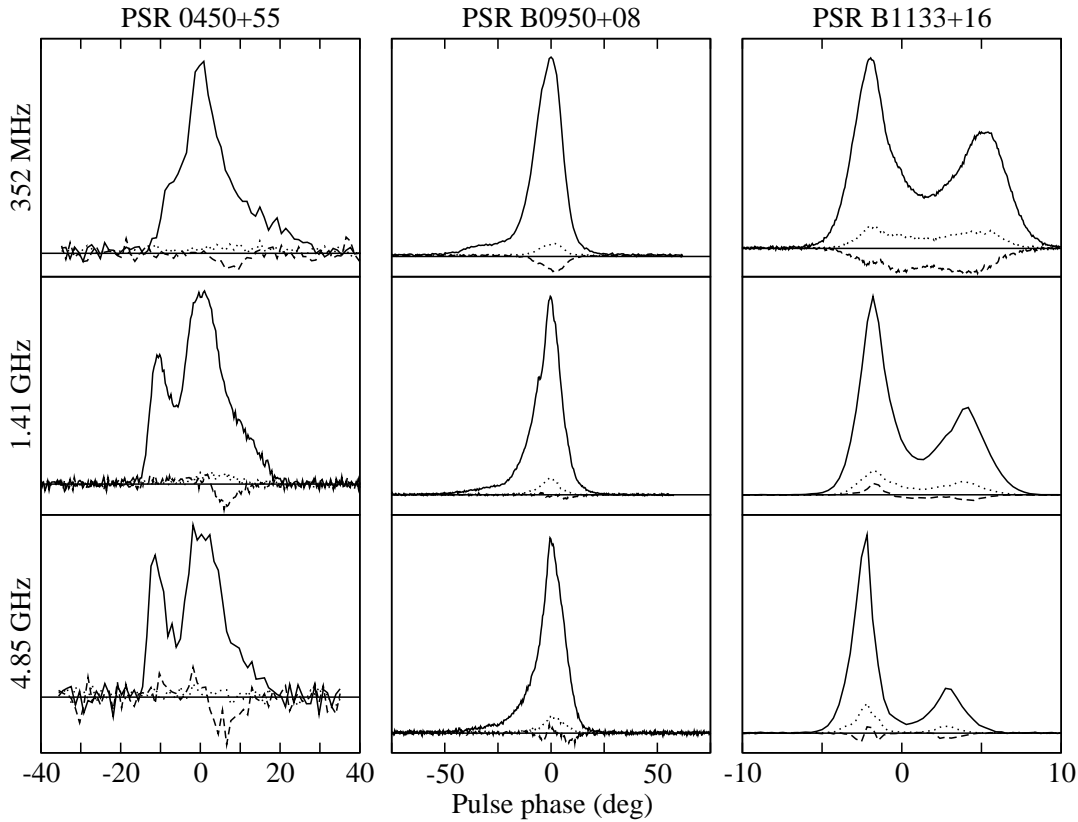}}
}
\caption{ Profiles of PSRs B0450+55, B0950+08 and B1133+16 observed at
three frequencies. The solid line represents the total power, the
dashed line represents $V$ and the dotted line $|V|$. For clarity, the
linear polarisation is not plotted.}
\end{figure*}
\begin{figure*}
\centerline{
\resizebox{\hsize}{!}{\includegraphics[angle=-90]{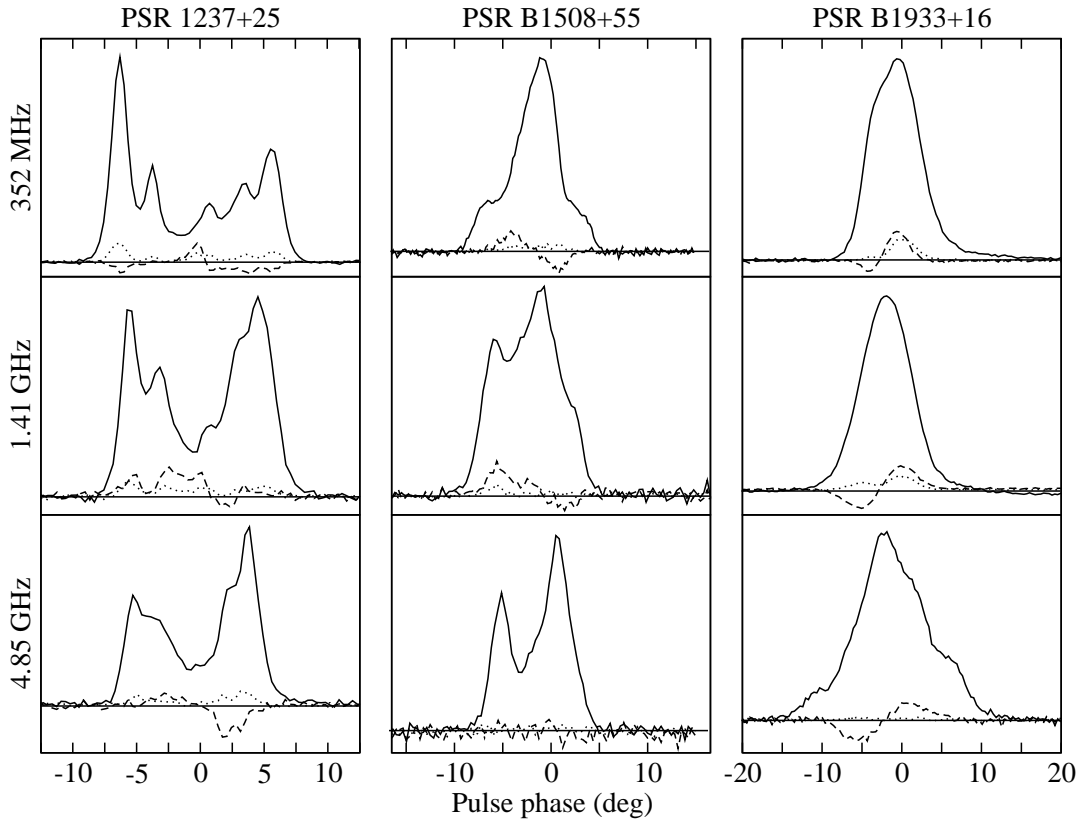}}
}
\caption{
Same as Fig. 1 for PSRs B1237+25, B1508+55 and B1933+16.
}
\end{figure*}
\begin{figure*}
\centerline{
\resizebox{\hsize}{!}{\includegraphics[angle=-90]{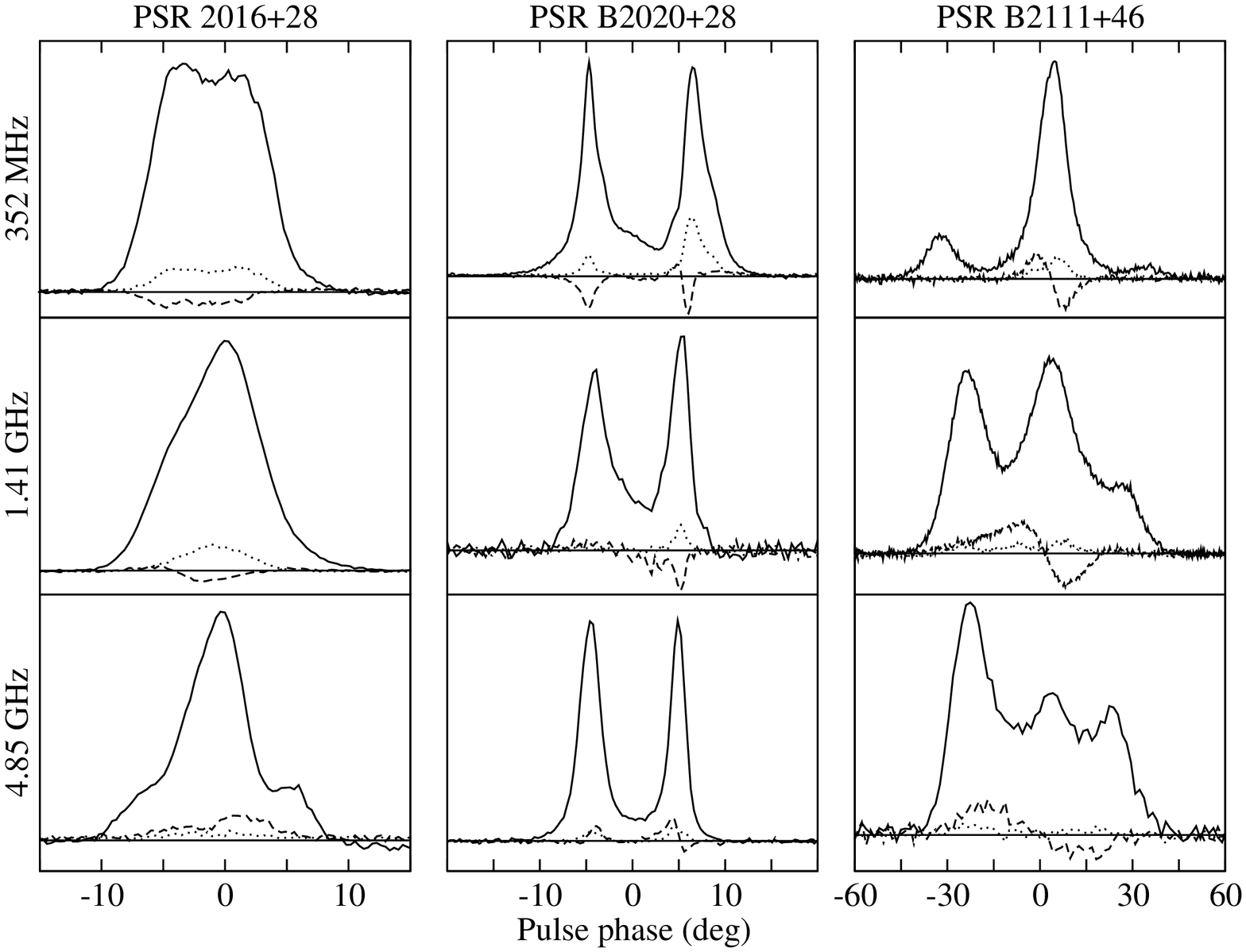}}
}
\caption{ Same as Fig. 1 for PSRs B2016+28, B2020+28 and B2111+46.  }
\end{figure*}

\subsection{Circular polarisation in core components}
PSRs B0450+55, B1237+25, B1508+55, B1933+16 and B2111+46 (in Figs. 1,
2 and 3) all have identifiable core components, demonstrating steep
position angle swings (Lyne \& Manchester 1988\nocite{lm88}) and, in
most cases, a swing in the $V$ profile from one to the other
handedness located at the same pulse phase. This can be seen most
clearly in PSRs B1933+16 and B2111+46. In both these pulsars, the
swing in the $V$ profile is evident at all three frequencies. The
profile of $|V|$ shows a minimum which coincides exactly with $V$
changing handedness. This is evidence that $V$ in the single pulses
not only has a mean of zero, but is also tightly distributed around
its mean, because a broad distribution would include pulses of high
S/N of either handedness which would inevitably increase $|V|$. This
indicates that the $V=0$ phase remains fixed in the single pulses. We
interpret this as direct evidence that the $V$ swing is a constant
feature of the single pulses.

We also notice that the swing in $V$ in the core component of both
these pulsars sweeps across a wider pulse phase range with
frequency. Apart from the two aforementioned pulsars, the $V$ and
$|V|$ signature described here can also be seen in all but the highest
frequency in PSR B1508+55.

\subsection{Circular polarisation in cone components}

PSRs B0950+08 and B1133+16 in Fig. 1 both consist of cone components
according to Lyne and Manchester (1988). The fact that the $V$
profiles at 1.41 and 4.85 GHz are very weak is in accordance with the
statement that $V$ is generally low in cone components
(e.g. Radhakrishnan \& Rankin 1990). However, as can be seen in
Fig. 1, $|V|$ is quite high in these components at these frequencies,
making it comparable to $V$ in the core components of other pulsars in
our sample. The mechanism, therefore, that is responsible for circular
polarisation in the single pulses appears to be almost equally
efficient between core and cone components.

\subsection{Frequency dependence of $V$ and $|V|$}

Apart from the observation that the cone components in PSRs B0950+08
and B1133+16 show high $|V|$, our data permit an investigation of $V$
and $|V|$ as a function of frequency.  At the lowest frequency, $V$ is
negative in both pulsars and the profile of $|V|$ is approximately a
mirror image of that of $V$. At 1.41 GHz, $V$ is significantly less
than at 352 MHz. The profile of $|V|$ however is similar to that at
352 MHz. This can be caused by $V$ in the single pulses being
symmetrically distributed around zero, albeit with cases of large
deviation from the mean. The fraction $|V|$/$I$ in the single pulses
is approximately the same between these two frequencies and it is the
symmetrical distribution of $V$ around zero in the single pulses, that
causes the $V$ profile to be almost zero. Moving further up in
frequency, the 4.85 GHz profile of $V$ has a more complex structure,
with a number of phases of transition between left and right
handedness. Despite this, the profile of $|V|$ remains the same even
at this frequency for both pulsars. The constance of the $|V|$ profile
with frequency regardless of the changes in $V$ constitutes evidence
that $|V|$, the flux density of the circularly polarised component of
the radiation, is a fundamental quantity of the emission process.

PSR B2016+28 in Fig. 3 shows the $|V|$ profile to be wider than the
$V$ profile at the two lowest frequencies. Especially at 1.41 GHz, at
either side of the $V$ component the $|V|$ component drops to zero
slower. In the leading edge of the component, it appears as thought
$V$ is slightly left handed, but despite the change in handedness, the
$|V|$ profile is approximately the same fraction of $I$ as at 352
MHz. This change of handedness appears to be part of a smooth change
with frequency, which is responsible for the 4.85 GHz profile being on
average slightly left handed across the pulse. It seems, therefore,
that this pulsar resembles PSRs B0950+08 and B1133+16. In fact,
Figs. 1, 2 and 3 show that $|V|$ is much less dependent on frequency
than $V$ itself. The 4.85 GHz data may be slightly harder to
interpret, because of a lack of S/N in the single pulses, but there is
no pulsar that behaves contradictory to the above statement at 4.85
GHz.

\subsection{The orthogonal polarisation mode phenomenon and $|V|$}

Both PSR B0950+08 and B1133+16 are pulsars that show OPM jumps in the
integrated polarisation profiles. The OPM phenomenon has been shown to
be more prominent at higher frequencies (Stinebring et al. 1984a, b,
Karastergiou et al. 2002\nocite{kkj+02}) and it is our opinion that
the $V$ profile changes seen in these two and other pulsars from our
sample, are OPM related. A full discussion on this is beyond the scope
of this letter, it is however part of the more detailed approach to
the behaviour of $|V|$ in Karastergiou et al. (2003b). The
orthogonal modes have been shown to be associated to a particular
handedness of circular polarisation (Cordes et al. 1978 \nocite{crb78}
Stinebring et al. 1984 a, b\nocite{scr+84}\nocite{scw+84}), however it
has been shown recently that this association becomes weaker at higher
frequencies for PSR B1133+16 (Karastergiou et
al. 2003\nocite{kjk03}). This results in a randomisation of the
handedness of the circular polarisation which effectively reduces the
mean $V$, but leads to a relatively high $|V|$. The relevance of the
OPM phenomenon can be also seen in PSR B2020+28 in Fig. 3, especially
at 352 MHz. In the trailing component of the pulse, $V$ sharply turns
from positive to negative and back to being slightly positive. These
sudden swings, which do not seem to affect the wider component of the
$|V|$ profile occur together with OPM jumps in the PA swing. It
therefore seems that also in this case the OPM phenomenon has the
effect of reducing the mean $V$ while $|V|$ remains largely
unaffected. This is also a good example of a $V$ swing without a $|V|$
minimum at the cross-over point, as is the case in the core
components. Obviously, in PSR B2020+28, at the pulse phase of the
change in the sign of the mean $V$, the single pulses will be somewhat
circularly polarised, with cases of either handedness and a mean of
zero.

\section{Conclusions}

We have demonstrated that accompanying the integrated profile of $V$
with the profile of $|V|$ provides important information on the
behaviour of circular polarisation in pulsar radio emission. Making
use of this simple quantity, which requires single pulse observations,
has immediately revealed new properties in a group of otherwise well
studied bright stars. We have shown that:
\begin{enumerate}
\item In the pulsars that exhibit a sign change signature in the core
components, a minimum in the $|V|$ profile is seen to coincide with
the point $V$ changes handedness. We consider this direct evidence
that the phase where the mean $V$ is 0, must have $V\approx0$ also in
the single pulses and, subsequently, that the swing signature is also
constant in the single pulses. Multiple frequency observations
demonstrate that this fact is true at all observed frequencies.
\item In a number of cases, the profile of $|V|$ resembles the profile
of $I$, regardless of the complicated structure seen in $V$. This is
also seen to be frequency independent. Complicated $V$ profiles are
seen in pulsars where the OPM phenomenon is observed, and, as such,
appear predominantly towards the higher frequencies.
\item In pulsars with cone components, we have shown that the very
weak $V$ profiles that may be due to the OPM phenomenon, show
comparatively high $|V|$. This demonstrates that the single pulses may
occasionally be highly circularly polarised, but average out to zero.
\end{enumerate}
Points (i) and (iii) indicate that the difference in terms of $V$
between core and cone components is related to how constant the $V$
features are in the single pulses. This is also in accordance with the
observation that core components are more stationary in phase with
respect to their cone counterparts that often tend to jitter. Such
jittering of a component which is partially left- and partially
right-hand circularly polarised will have the effect of reducing $V$,
while $|V|$ should survive. All three points mentioned above warrant
further investigation using $|V|$ to determine more accurately the
nature of $V$ in pulsars. They convincingly suggest once again, that
in the effort for progress in solving the pulsar emission mechanism
problem, single pulse observations carry vital information that should
not be underestimated and lost due to integration.\\
\noindent{\bf Acknowledgments}~~We thank B.~Stappers for help with
the observations. SJ thanks R.~Wielebinski for hospitality during his
visit to the MPIfR.

\label{lastpage}
\bibliographystyle{mn2e}
\bibliography{journals,modrefs,psrrefs}
\end{document}